\documentclass[10pt,aps,prb,amsmath,tightenlines,notitlepage]{revtex4}
\usepackage{color}
\usepackage{amsmath}
\usepackage{graphicx}

\makeatletter
\@ifundefined{textcolor}{}
{%
 \definecolor{BLACK}{gray}{0}
 \definecolor{WHITE}{gray}{1}
 \definecolor{RED}{rgb}{1,0,0}
 \definecolor{GREEN}{rgb}{0,1,0}
 \definecolor{BLUE}{rgb}{0,0,1}
 \definecolor{CYAN}{cmyk}{1,0,0,0}
 \definecolor{MAGENTA}{cmyk}{0,1,0,0}
 \definecolor{YELLOW}{cmyk}{0,0,1,0}
}

\usepackage{xcolor}
\usepackage{multirow}
\usepackage{textcomp}
\usepackage{wrapfig}
\makeatother

\begin{document}
\title{Use of multigrids to reduce the cost of performing interpolative separable density fitting}
\author{Kori E. Smyser}
\affiliation{Department of Chemistry, University of Colorado, Boulder, CO 80302,
USA}
\author{Alec White}
\affiliation{Quantum Simulation Technologies, Inc., Boston, 02135, United States}
\author{Sandeep Sharma}
\email{sanshar@gmail.com}

\affiliation{Department of Chemistry, University of Colorado, Boulder, CO 80302,
USA}
\begin{abstract}
In this article, we present an interpolative separable density fitting (ISDF) based algorithm to calculate exact exchange in periodic mean field calculations. In the past, decomposing the two-electron integrals into tensor hypercontraction (THC) form using ISDF was the most expensive step of the entire mean field calculation. Here we show that by using a multigrid-ISDF algorithm both the memory and the CPU cost of this step can be reduced. The CPU cost is brought down from cubic scaling to quadratic scaling with a low computational prefactor which reduces the cost by almost two orders of magnitude. Thus, in the new algorithm, the cost of performing ISDF is largely negligible compared to other steps. Along with the CPU cost, the memory cost of storing the factorized two-electron integrals is also reduced by a factor of up to 35. With the current algorithm, we can perform Hartree-Fock calculations on a Diamond supercell containing more than 17,000 basis functions and more than 1,500 electrons on a single node with no disk usage. For this calculation, the cost of constructing the exchange matrix is only a factor of four slower than the cost of diagonalizing the Fock matrix. Augmenting our approach with linear scaling algorithms can further speed up the calculations.
\end{abstract}
\maketitle

\section{Introduction}
Much of modern {\it ab initio} computational chemistry and materials science is based on Kohn-Sham (KS) density functional theory (DFT)\cite{kohn1965self}. The inclusion of exact Hartree-Fock (HF) exchange within this framework\cite{b3} has been instrumental to the success of DFT for molecular systems to the point that almost all modern molecular calculations rely on these ``hybrid'' density functionals\cite{mardirossian2017thirty}. Hybrid functionals can outperform their semi-local counterparts for some properties of periodic solids\cite{Heyd2005,paier2006screened,Finazzi2008,Hai2011,Basera2019,kovacic2020photocatalytic}, but, in contrast to molecular calculations, the cost of evaluating the non-local exchange contribution may be prohibitive.

Methods for the efficient evaluation of exact exchange are well-developed in the context of molecular calculations. Such calculations typically use a relatively small set of small, local basis functions such that the ratio of basis functions to electrons, $N/n$, is often less than ten. Computing every element of the fourth-order tensor of electron repulsion integrals, which one might naively expect to be necessary for both the Coulomb and exchange contributions, would scale like $O(N^4)$. However, the locality of the basis functions implies that there are asymptotically only a linear number of {\it significant} basis function pairs, which means that the Coulomb and exchange contributions can be computed in $O(N^2)$ time\citep{Almlof1982}. The scaling of the Coulomb contribution can be reduced to linearithmic, $O(N\ln N)$, using the multipole expansion and fast multipole method\cite{White1994,challacombe1996fast,Kudin1998}. The exchange contribution can also be computed in asymptotically linear time for non-metallic systems by leveraging locality in the density matrix\citep{Challacombe1997,Ochsenfeld1998,Goedecker1999,Ko2020}. For most practical molecular calculations, these asymptotically linear methods come with a large prefactor, and it is preferable to reduce the cost of higher-scaling algorithms with tensor factorization. The resolution of the identity (RI) method is one such tensor factorization technique that is commonly applied to both Coulomb and exchange contributions with the RI-J\citep{Sierka2003,Sodt2006} and RI-K\citep{Polly2004,Sodt2008,Manzer2015,Manzer2015a} algorithms respectively. These RI methods are also called ``density fitting;'' and Dunlap showed how a ``robust'' fit can be used to make the error in the fitted two-electron integral quadratic in the error for basis function pairs.\citep{Dunlap2000,Dunlap2000a,Hollman2017} RI approaches usually rely on predetermined, atom-centered basis sets of fitting functions. Circumventing this requirement, local-RI using numerical basis\cite{Ihrig_2015,kokott2024efficient} and Cholesky decomposition approach yields a factorization of the same form without the need for pre-optimized fitting basis sets.\citep{Beebe1977,Aquilante2007} The pseudospectral (PS) method is an alternative factorization that uses a partial real-space quadrature to factorize the two-electron integrals\citep{Friesner1985}, and the chain of spheres algorithm for exchange (COSX)\citep{Neese2009} is a commonly used implementation of the PS idea. In recent years, the tensor hypercontraction (THC) method of Martinez and coworkers took the idea of tensor factorization to the logical limit \citep{Hohenstein2012,Parrish2012,Hohenstein2012a}. The THC method factorizes the four-index tensor of two-electron integrals into a product of two-index tensors---a drastic factorization. But obtaining an accurate THC factorization is generally difficult so initial applications to correlated methods are limited. However, the ``interpolative separable density fitting'' (ISDF) method\citep{Lu2015} can provide a factorization of THC form with only cubic, $O(N^3)$, scaling, and it has since been used in various algorithms for exact exchange\citep{Hu2017,Dong2018,Lee2020,Sharma2022,rettig2023even,zhang2024machine}.

On the other hand, calculations on periodic solids often use a large basis set of $N_g$ plane waves. In these calculations, the action of the Coulomb operator on just the occupied space is determined by solving $n$ Poisson equations, which leads to quadratic scaling, $O(n N_g \ln N_g)$, or linear scaling when using translational ($k$-point) symmetry. Unfortunately, the action of the exchange operator on the occupied space is cubic, $O(n^2 N_g \ln N_g)$, or quadratic with $k$-point symmetry. So, in a plane-wave basis set, the exchange contribution is higher scaling than the Coulomb part, and there have been many efforts to reduce this cost. Linear scaling methods have been developed for both the Coulomb and exchange\citep{Goedecker1999,Bowler2012}. As in the molecular case, traditional linear scaling exchange algorithms rely on locality in the density matrix for insulating systems\citep{Wu2009}. An exception is stochastic density functional theory (sDFT), which can reduce the prefactor and scaling of the exchange calculation by using the stochastic resolution of identity method\citep{Baer2013,Neuhauser2016,bradbury2023}. For typical calculations, asymptotically linear scaling methods are not practical, and methods to improve the efficiency without addressing the scaling can result in useful speedups\citep{Todorova2006}. Additional examples include the adaptively compressed exchange (ACE) method\citep{Lin2016} and the auxiliary density matrix method\citep{Guidon2010}. Methods that use ISDF-THC for exact exchange in solids\citep{Hu2017,Dong2018,Sharma2022,rettig2023even,zhang2024machine}, including the method presented in this work, fall into this category.

Since its introduction by Lu and Ying in 2015,\citep{Lu2015} ISDF has been quickly adopted to speed up the exchange calculations in codes that use Gaussian orbitals\citep{Sharma2022,rettig2023even}, numerical atomic orbitals\cite{qin2020interpolative}, and plane wave basis sets\citep{Hu2017,Dong2018,zhang2024machine}. Although the computational scaling of performing ISDF is cubic with the system size (the same as the ultimate computation of the exchange matrix),
the computational prefactor is high, making it one of the most expensive steps of the entire calculation. Furthermore, the memory requirements are high,  which limits its applicability to small systems unless massively parallel computers are used. Recently, it has been realized that the memory cost can be reduced if one performs interpolative decomposition of the occupied molecular orbitals rather than the entire atomic orbital basis set\cite{rettig2023even}. This does reduce the memory requirement but this comes at an additional cost of having to perform this decomposition at every self-consistent field (SCF) iteration. There is also additional computational overhead related to constructing the exchange matrix (for details we refer the reader to Ref.~\onlinecite{rettig2023even}). 

In this work, we perform the interpolative decomposition on the atomic integrals, which is only done once, and simultaneously reduce the computational and memory cost of performing this step using ideas from so-called ``multigrid'' approaches\cite{Lippert1999,Vandevondele2005,laino2005efficient,beck20095,del2012second,Kuhne2020}. As we will show in the results, the computational cost of this step is no longer the leading cost of the algorithm and we can perform calculations on systems with $>10000$ basis functions on a single node without running out of memory.  Along with reduced memory requirements, we also show that the cost of performing ISDF calculations is nearly eliminated. The key ingredient of the algorithm is to use multiple local grids of varying resolutions, each of which supports only a subset of atomic orbitals (Fig.~\ref{fgr:Grids}). This idea has been used in the past to significantly speed up the calculation of the Coulomb operator in CP2K\cite{Vandevondele2005,Kuhne2020}, and here we extend this approach to accelerate exchange evaluation within the ISDF-THC framework. 

In the rest of the paper, we will focus on periodic calculations with Gaussian basis functions in the presence of Pseudopotentials, although the ideas can be extended to mixed Gaussian/plane-wave basis and all-electron calculations. We will begin the paper by recalling how the interpolative decomposition is typically performed to obtain integrals in the THC form. This will be followed by our updated algorithm that shows how the memory and CPU cost of performing this step can be significantly reduced. We end the paper with some results and prospects for future work.

\section{THEORY}
\begin{table}[h]
\caption{Notation used in paper.\label{tbl:Notation}}
\centering
\begin{tabular}{l l} 
\hline
$N_g$ & Number of grid points\\
$N_{\xi}$ & Number of fitting functions\\
$N$ & Number of basis functions\\
$n$ & Number of electrons\\
 $n_{atom}$&Number of atoms\\
$\mu, \nu, \cdots$ & Indices of atomic orbitals\\
$i,j,\cdots$ & Indices of the occupied molecular orbitals\\
$p,q,\cdots$ & Indices of any molecular orbital\\
\hline
\end{tabular}
\end{table}

A major bottleneck in a simple implementation of hybrid-DFT calculations is the need to evaluate the two-electron integrals,
\[
(\mu\nu|\lambda\sigma)=\int\int\mu(r_{1})\nu(r_{1})\frac{1}{r_{12}}\lambda(r_{2})\sigma(r_{2})dr_{1}dr_{2}
\]
Instead of calculating the entire four-index quantity one can decompose it into a product of several two-index quantities. We begin by noting that the product of the orbitals $\mu(\mathbf{R})\nu(\mathbf{R})=(\mu\nu|\mathbf{R})$ can be viewed as a matrix with two indices, the first index being a composite index consisting of a pair of orbitals, $\mu\nu$, and the second being a set of suitably chosen grid points with a sufficiently high density, $\mathbf{R}$. This matrix $(\mu\nu|\mathbf{R})$ is low-rank and can be decomposed as,
\[
(\mu\nu|\mathbf{R})=\sum_{\xi}(\mu\nu|\xi)\xi(\mathbf{R})
\]
where the size of the index $\xi$ is smaller than both the square of the number of basis functions ($N$) and the number of grid points ($N_g$). An optimal decomposition that minimizes the Frobenius norm of the error is given by SVD. However, with SVD we lose the separability of the original matrix (note that while $(\mu\nu|\mathbf{R})=\mu(\mathbf{R})\nu(\mathbf{R})$, $(\mu\nu|\xi)\neq\mu(\xi)\nu(\xi)$). One can instead perform interpolative decomposition that ensures that the indices $\xi$ in the matrix $(\mu\nu|\xi)$ are just a subset of grid points $\mathbf{R}$. The disadvantage of interpolative decomposition is that one does not have an optimal algorithm to find it and the Frobeius norm of the error is guaranteed to be greater than that from SVD, but the separability of the resulting matrix is retained, i.e. $(\mu\nu|\xi)=\mu(\xi)\nu(\xi)$, which more than makes up for the shortcomings.

After having performed the interpolative decomposition of the orbital products one can then write the two-electron integrals as,
\[
(\mu\nu|\lambda\sigma)=\sum_{\xi\xi'}\mu(\xi)\nu(\xi)(\xi|\xi')\lambda(\xi')\sigma(\xi')
\]
 where the matrix $(\xi|\xi')$ can be evaluated numerically as,
\begin{equation}
    (\xi|\xi')=\int\int\xi(r_{1})\frac{1}{r_{12}}\xi'(r_{2})dr_{1}dr_{2}
    \label{eqn:(xi|xi')}
\end{equation}
 using fast Fourier transform (FFT). 

Typically, THC requires more fitting functions $\xi(r)$ than RI---while the error in the two-electron integrals is
quadratic in the RI fitting error, it is linear in the THC error. One can use robust tensor hypercontraction, also known as the robust pseudospectral (rPS) method, to make the error from two-electron integrals quadratic in the fitting errors (similar to RI) and use fewer functions, as in RI.  rPS is known to produce non-positive definite two-electron integrals that can cause variational collapse of the SCF cycles\cite{Merlot2013,Wirz2017}. But we have never seen this in our previous work because we never use rPS to evaluate the Coulomb matrix---only the exchange matrix, which is itself negative definite. 

\subsection{Interpolative decomposition}\label{sec:ISDF_review}

As mentioned in the previous section, one needs to perform an interpolative decomposition of the two-electron integrals. The most common way of doing this is to perform pivoted-QR decomposition of the $(\mu\nu|\mathbf{R})$ matrix. A simple algorithm would lead to a computational cost of $O(N^{4})$, making the entire algorithm prohibitively expensive. Lu and Ying\citep{Lu2015} in their original paper introduced a randomized algorithm where one first obtains two random matrices $G^{1}$ and $G^{2}$ of size $N\times p$ each, with $p=\sqrt{N_{\xi}}+\delta$ orthogonal columns, where $N_{\xi}$ is the number of THC functions and $\delta $ is a small number usually around 5\citep{Halko2011,Liberty2007}. A randomized density matrix is constructed from these matrices according to,
\[
\rho_{mn,R}=\left(\sum_{\mu}G_{\mu m}^{1}\mu(\mathbf{R})\right)\left(\sum_{\nu}G_{\nu n}^{2}\nu(\mathbf{R})\right)
\]
One can then perform a pivoted-QR decomposition on the matrix $\rho_{mn,R}$ to obtain the pivots. The pivots from the randomized matrix will be of similar quality to those obtained from the full matrix $(\mu\nu|\mathbf{R})$ as long as its singular values decay sufficiently quickly, as they do here. The overall cost of the randomized algorithm is $O(N^{3})$ which is a significant improvement over the deterministic algorithm. 

Matthew suggested\cite{Devin2020} that one can improve the efficiency of the algorithm by first forming a matrix 
\begin{equation}
   M(\mathbf{R},\mathbf{R}')=\sum_{\mu\nu}(\mu\nu|\mathbf{R})(\mu\nu|\mathbf{R}')
   \label{eqn:M(R,R)}
\end{equation}
and then perform a pivoted-Cholesky decomposition on it to obtain the pivots $\xi$. The methods give the same pivot points (when randomization is not introduced), and pivoted Cholesky is typically significantly faster than pivoted QR. This algorithm is extremely efficient especially if the matrix $M$ can be stored in memory. Later we will show that for our purposes these matrices are indeed small enough to be stored in memory. It is also worth mentioning that a third approach called centroid-Voronoi-tesselation (CVT) is also widely used in this context\cite{Hu2017,Lee2020}. 

Having obtained the pivot points $\xi$, a least square algorithm obtains the functions $\xi(\mathbf{R})$ that minimize the error,
\begin{equation}
\min_{\xi(\mathbf{R})}\left|(\mu\nu|\mathbf{R})-\sum_\xi(\mu\nu|\xi)\xi(\mathbf{R})\right|
   \label{eqn:lsqr}
\end{equation}
This can be done with an $O(NN_\xi N_g + N_\xi^3 + N_\xi^2 N_g)$ cost, dominated by $O(N_\xi^2 N_g)$. As mentioned in the introduction, this algorithm has a rather steep memory requirement because one has to store the fitting functions $\xi(\mathbf{R})$ at the cost of $N_{g}\times N_{\xi}$. The value of $N_{g}$ can become significant even if there is a single sharp function in the basis set.

There are a few ways of overcoming the high cost of ISDF calculation:
\begin{enumerate}
\item In a previous publication,\cite{Sharma2022} we have shown that one can reduce both the memory and CPU cost of ISDF by using a robust fitting procedure, which reduces the number of ISDF functions needed to get an accurate result by about a factor of two (for instance, compare THC and rPS in Figure~\ref{fgr:bar_chart}). Although the cost of ISDF is reduced, it remains the dominant cost of the calculation.
\item The cost of doing ISDF can be eliminated by not doing ISDF but instead by solely relying on FFT and using the occ-RI (occ refers to occupied orbitals) trick of Manzer {\it et al.}\cite{Manzer2015a} occ-RI relies on the fact that the value and gradient of the DFT energy can be obtained simply by knowing the occupied-virtual block of the exchange matrix $K_{ip}$, which is given by
    \begin{align}
        K_{ip} = \sum_p\int\int \phi_i(r_1) \phi_j(r_2)\frac{1}{r_{12}}\phi_j(r_1) \phi_p(r_2) dr_1 dr_2
    \end{align}
where we have assumed that the orbitals are real. 
   
If all the orbitals are representable on an FFT grid of size $N_g$ then this entire matrix can be evaluated by performing $n^2$ Poisson solves, and matrix multiplications with the cost equal to $O(n^2 N_g \ln(N_g))$ and $O(nNN_g)$, respectively. Out of the two steps, we find that the cost of Poisson solves $O(n^2 N_g \ln(N_g))$ dominates. Because the ISDF calculation is not used,  the memory requirement for storing the fitting functions  $\xi(\mathbf{R})$ ($N_{g}\times N_{\xi}$) is eliminated. 

\item Instead of performing ISDF on the products of atomic orbitals once at the start of the calculation, one can perform a new ISDF calculation on the product of molecular orbitals at each SCF iteration.
The two dominant costs of this algorithm are the same as that of AO-based ISDF---$O(N_\xi^2 N_g)$ for matrix multiplications and $O(N_\xi N_g\ln(N_g))$ for $N_\xi$ Poisson solves. The potential advantage is that the $N_\xi$ required is independent of the basis set, it only depends on the number of electrons, and it is expected to be smaller than the $N_\xi$ from ISDF on atomic orbital pairs. The disadvantage is that one has to perform an ISDF at each SCF iteration and ISDF remains the dominant cost of the calculation. This approach, of performing ISDF on the molecular orbitals, was first pointed out by Hu et al.\cite{Hu2017} It was recently extended to use with Gaussian basis sets and $k$-point sampling by Rettig et al.\cite{rettig2023even} where they pointed out a few terms that were missing in the gradient of the exchange energy in \citep{Hu2017}.
\end{enumerate}

For $\Gamma$-point calculations we expect the CPU cost of the FFT-based approach and rPS to be lower than that of the MO-based ISDF calculations. The memory cost of the FFT-based approach is superior to the other two because one does not have to store the ISDF fitting functions.

In this work, we introduce a third approach that relies on the use of multiple grids of varying resolutions. Our algorithm uses both a single-shot ISDF by using pivoted Cholesky on atomic orbitals and an iterative FFT-based Poisson solution at each SCF cycle. The ISDF is only performed for products of atomic orbitals where at least one of the orbitals is sharp and the FFT is only used to solve the Poisson equation for products where both orbitals are diffuse, as described in more detail below.

\subsection{Using multiple grids for exchange}\label{sec:multigrid_exchange}
In this section we describe the basic idea of our multigrid algorithm for calculating exchange and go into more technical details in the next section. 
We begin with an uncontracted Gaussian-type orbital (GTO) basis and partition it into two sets. The first set contains sharp Gaussian basis functions with large exponents and the second set contains diffuse basis functions with small exponents. The product of sharp-sharp and sharp-diffuse atomic orbitals are approximated using ISDF on a grid centered around the atom on which the sharp function is centered (see Figure~\ref{fgr:Grids}). Because the functions are sharp here, they require the grid to have a high point density, though for the same reason they do not span the full unit cell. Notice that the sharp functions with large exponents decay rapidly and their product with any other function is only expected to be non-zero in the local spatial region where it is itself non-zero.
The rest of the products between diffuse-diffuse functions are treated using the iterative FFT approach outlined in the second bullet point of Section~\ref{sec:ISDF_review}. This does not use ISDF but is solely based on Poisson solves using FFT.  Our algorithm can reduce both the memory and CPU cost compared to usual ISDF-based calculations because: 
\begin{enumerate}
\item We only need to perform ISDF calculations once, before the SCF iterations, instead of at each SCF iteration. We perform $n_{atom}$ (the number of atoms) independent local ISDF calculations and each calculation is cheap. The cost of each local ISDF calculation scales linearly with the size of the system. After the ISDF fitting functions are obtained one has to construct the two-center Coulomb matrix $(\xi|\xi')$ which scales quadratically with the size of the system, making the entire ISDF calculation quadratic in system size. In this algorithm, we do not need to use the randomized algorithm and the overall cost of ISDF is negligible. 
\item The memory cost of our algorithm is largely independent of the size of the basis set and the overall memory requirement is quite low. This is because all of the diffuse-diffuse products, which represent the largest fraction of the non-negligible basis set pairs, are treated using an FFT grid that is independent of the basis set size. A key point is that the number of grid points needed to represent diffuse functions can be fairly small and thus the Poisson solves are cheap. Furthermore, for these pairs of basis functions, we do not store ISDF functions. 
\end{enumerate}

Because we use multiple grids of different resolutions, we have called our method ``multigrid'' inspired by the approach of the same name used to speed up the Coulomb matrix formation\cite{Lippert1999,Vandevondele2005,laino2005efficient,beck20095,del2012second,Kuhne2020}. Although currently we only employ grids of two resolutions, one for sharp and one for diffuse functions, this approach is readily extended to include a larger number of grids. This can become important when all-electron calculations are performed without the use of Pseudopotentials, where a larger range of resolutions is needed due to the presence of extremely sharp basis functions. The approach here also has similarities to that of F{\"u}sti-Moln{\'a}r and Pulay\cite{fusti2002accurate, fusti2002FFT} in which the basis functions are partitioned into sharp and diffuse and different approaches are used to evaluate the contributions from various pairs. Recently, a similar approach has been used in the context of stochastic density functional theory where contributions of sharp functions are evaluated exactly while for diffuse functions stochastic resolution of identity is used\cite{bradbury2023}. 

In the next section, we describe in more detail how the various grids are formed and how the entire exchange matrix is calculated.

\section{Computational details}
\begin{figure}[ht]
\centering
\includegraphics{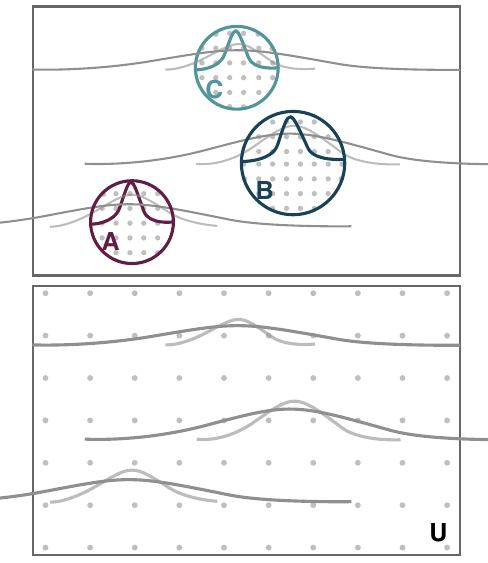}
\caption{Example of the grid structure used for building the exchange matrix. The most dense grid is only defined on spherical, atom-centered regions (colored circles, $A$, $B$, $C$; top). The volume of each atom-centered grid is determined by a cutoff radius beyond which the local GTOs ($L$) with exponents larger than $\alpha_{\min}$ (colored lines) are expected to go to zero (see Equation~\ref{eqn:rcut}). A sparse universal grid ($U$) spans the full supercell and supports all global GTOs ($G$) but is only required by exponents smaller than $\alpha_{\min}$. }\label{fgr:Grids}
\end{figure}

To illustrate the algorithm let us imagine we have an atom, $A$, with uncontracted GTOs with exponents $\alpha_{1},\alpha_{2},\cdots$ in decreasing order of magnitude. A user-defined value, $\alpha_{\min}$, divides the GTOs into sharp functions that have exponents larger than $\alpha_{\min}$ and diffuse functions with exponents smaller than $\alpha_{\min}$. Although the sharp functions require a high-resolution grid, the spatial extent of this grid is relatively small because the function decays rapidly in real space. The real space and Fourier space representation of an $s$-type function of exponent $\alpha$ are $\exp(-\alpha r^{2})$ and $\exp(-G^{2}/(4\alpha))$ up to a multiplicative factor. If we want to represent the functions in real space up to an accuracy of $\varepsilon_r$ then we have a local atom-centered grid of radius,
\begin{equation}
    \begin{aligned}
            r_{\max} =\sqrt{\frac{-\ln(\varepsilon_{r})}{\alpha_{\min}}}
    \end{aligned}
    \label{eqn:rcut}
\end{equation}
All exponents greater than $\alpha_{\min}$ are supported by local grids with grid points $\mathbf{R}_A, \mathbf{R}_B,\cdots$ respectively, centered on atoms $A, B, \cdots$, respectively (see the upper panel of Fig~\ref{fgr:Grids}). The remaining functions are represented on a sparse grid of lower resolution that spans the entire unit cell ($U$ in the lower panel of Fig~\ref{fgr:Grids}). The sparse universal grid is truncated in the Fourier space with a wave number $G_{U,\max}$ such that,
\begin{equation}
    \begin{aligned}
            G_{U,\max} & =\sqrt{-4\alpha\ln(\varepsilon_{K})}\\
    \end{aligned}
    \label{eqn:Ggrid}
\end{equation}
where $\varepsilon_K$ is a threshold that one can decrease continuously to increase the overall accuracy of the calculation. 

For all systems considered in this article, we have found that using $\varepsilon_r=10^{-5}$ gives an overall error that will be below 50 $\mu$ Ha per atom.
The optimal value of $\alpha_{\min}$ from a computational cost point of view can be system-dependent. There is a trade-off between the cost of the exchange evaluation and the memory requirement for storing the functions on the dense grid. A higher $\alpha_{min}$ speeds up the ISDF calculation because fewer functions are considered sharp but it also increases the memory cost for storing $\mu(\mathbf{R}_U)$ because the universal grid $U$ requires more points. We have found that an $\alpha_{min}$ value of 2.8 $\text{Bohr}^{-2}$ is a reasonable choice for the systems studied here.

Now one performs a local interpolative decomposition on each atom-centered grid $A$ such that the equality,
\begin{equation}
   \mu^L_{A}(\mathbf{R}_{A})\nu^G_{A}(\mathbf{R}_{A})\approx\sum_{\xi_{A}}\mu^L_{A}(\xi_{A})\nu^G_{A}(\xi_{A})(\xi_{A}|\mathbf{R}_{A})
   \label{eqn:mg product density}
\end{equation}
is satisfied up to sufficient accuracy, determined by an ISDF threshold $\varepsilon_{\mathrm{ISDF}}$. In the equation, the subscripts indicate the grid centered on atom $A$ and the superscript $L$ stands for local, indicating all sharp functions that are atom-$A$ centered. $\nu^G_{A}$ are all functions (superscript $G$ stands for global) that have a non-zero value on the grid around atom $A$, these include both the sharp and diffuse functions on atom A and also functions on other atoms. $\xi_A$ are ISDF fitting points (functions) on the grid around $A$.
At this point, it is also useful to define the set of functions $\mu^N_A$ that are present in the global list ($G$) but are not in the set ($L$), i.e. functions that have a non-zero value on at least one of the grid points $\mathbf{R}_A$ of local grid centered on $A$ but are not one of the sharp functions centered on atom $A$.

\begin{table}
    \centering
    \begin{tabular}{ll}
    \hline
        Symbol & Meaning \\
    \hline
    $\alpha_{\min}$ & Gaussians with exponents greater than $\alpha_{\min}$ are sharp and others are diffuse\\
      $\varepsilon_r$   & Threshold that determines the extent of the local grid (see Equation \ref{eqn:rcut} and upper panel of Figure~\ref{fgr:Grids})\\
      $\varepsilon_{ISDF}$ & Threshold used during pivoted-Cholesky decomposition that determines the number of ISDF functions\\
      $\varepsilon_K$ & Determines the number of grid points that make up the universal sparse grid (see Equation \ref{eqn:Ggrid})\\
      \hline
    \end{tabular}
    \caption{The accuracy of the exchange calculation is determined by four different thresholds as outlined above. For all the calculations in the paper we fix $\varepsilon_r$ to be $10^{-5}$ and $\alpha_{\min}$ is chosen to be 2.8 $\mathrm{Bohr}^{-2}$. The parameters $\varepsilon_K$ and $\varepsilon_{ISDF}$ are varied to obtain the desired accuracy.}
    \label{tab:ISDF-thresh}
\end{table}

For the dense atom-centered grids, the ISDF fitting is done using pivoted Cholesky without randomization because, unlike on a full grid, it is inexpensive to build the product density matrix $M(\mathbf{R}_{A},\mathbf{R}_{A}')$ (Eq.~\ref{eqn:M(R,R)}) on local grids. After selecting the points using pivoted Cholesky we use a least square minimization (Eq.~\ref{eqn:lsqr}) to obtain the fitting functions $\xi_{A}(\mathbf{R}_{A})$. These functions are fully supported on the local grids and are relatively inexpensive to store in memory. This local ISDF procedure is carried out for each local grid and the number of fitting functions chosen on each grid is controlled by the user-specified tolerance $\varepsilon_{ISDF}$. The cost of performing the local ISDF calculation for each atom-centered grid is system size-independent.

Once the fitting functions are formed we calculate the two-centered Coulomb integral $(\xi_A|\xi_B)$ for all pairs of atom-centered ISDF functions and $(\xi_A|\mathbf{R}_U)$, between ISDF functions on atom centered grids and each grid point on the universal grid. The cost of constructing these matrices is equal to $O(N_{\xi,L} N_g\ln(N_g)) + O(N_{\xi,L}^2)$, where $N_{\xi,L}$ are the total number of ISDF functions from all atom-centered grids. The first term comes from having to perform an FFT for each local ISDF fitting function and the second term comes from evaluating the matrix $(\xi_A|\xi_B)$. Notice that because each ISDF function is local the cost of this matrix evaluation is only quadratic as opposed to cubic with fully non-local ISDF functions. Finally, the calculation of the matrix $(\xi_A|\mathbf{R}_U)$ only requires the potential due to the fitting function on the universal grids and does not require any matrix multiplications. One can readily evaluate the entire matrix $(\mathbf{R}_U|\mathbf{R}_U')$ using a single FFT, using the fact that this is a circulant matrix. In our algorithm, we avoid storing this matrix and use FFT to calculate the exchange matrix. 

The analysis here shows that the entire cost of the ISDF calculation is quadratic in the system size which is lower than the cubic cost of forming the exchange matrix. Therefore, the overhead of performing these ISDF calculations is negligible. Table~\ref{tab:ISDF-thresh} summarizes the various thresholds used to construct the exchange matrix. 

\subsection{Building the exchange matrix}
\begin{figure}[ht]
\centering
\includegraphics{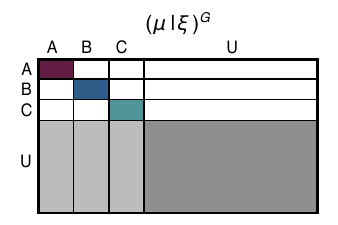}
\includegraphics{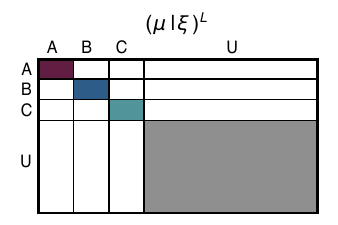}
\includegraphics{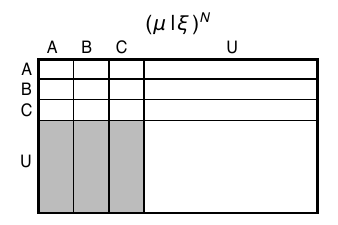}\\
\caption{Graphical representations of the matrices $(\mu|\xi)^{G}$, $(\mu|\xi)^{L}$, and $(\mu|\xi)^{N}$. The colors here follow from Figure~\ref{fgr:Grids} and represent sets of atom-centered GTOs that are local to atom-centered grids. The white blocks correspond to zeros in the matrix and the shaded blocks represent non-zero entries. The columns of the matrices correspond to ISDF functions for atom-centered grids ($A$, $B$, $C$) or all the grid points on the universal grid $U$. 
}
\label{fgr:(mu|xi)}
\end{figure}

In the multigrid approach, the expressions of the two-electron integrals have the THC form, however, there are now four terms,
\begin{align}
    (\mu\nu|\lambda\sigma) &= \sum_{\xi\xi'}(\mu|\xi)^L(\nu|\xi)^G(\xi|\xi')(\lambda|\xi')^G(\sigma|\xi')^L + \sum_{\xi\xi'}(\mu|\xi)^N(\nu|\xi)^L(\xi|\xi')(\lambda|\xi')^L(\sigma|\xi')^N \nonumber\\
    &=\sum_{\xi\xi'}(\mu|\xi)^L(\nu|\xi)^G(\xi|\xi')(\lambda|\xi')^L(\sigma|\xi')^N + \sum_{\xi\xi'}(\mu|\xi)^N(\nu|\xi)^L(\xi|\xi')(\lambda|\xi')^G(\sigma|\xi')^L \label{eq:ints}
\end{align}
The various matrices with superscripts $L$, $N$, and $G$ are described in Figure~\ref{fgr:(mu|xi)}. The four terms arise to ensure that products of basis functions are only evaluated on the appropriate grid. For example, there are diffuse functions that are non-zero on local grids of various atoms and also on the diffuse grid. The equation ensures that the product of diffuse functions is only evaluated on the most sparse grid; on the sharp grids, only products of sharp-sharp and sharp-diffuse are evaluated. 

We use occ-RI and only the occupied-virtual part of the exchange matrix is constructed using the two-electron integrals in Eq.~\ref{eq:ints}. We would like to reemphasize that in Eq.~\ref{eq:ints} one needs the two-electron integrals $(\mathbf{R}_U|\mathbf{R}_U')$ because all points on the sparse universal grids are included as ISDF points. However, this matrix is never stored and we rely on the fact that this matrix is diagonal in the Fourier space and the action of this matrix on any function is evaluated using FFTs. 

It is worth pointing out that one can potentially speed up the calculations by utilizing the block structure of the various matrices. However, in our current work, the entire code is implemented in Python and we find that implementing block multiplication by using for loops incurs an overhead that nullifies any benefit of reducing the computations. In a future publication, this can be remedied by implementing some of these matrix multiplications in an optimized C-code.

\section{Results and discussion}
The algorithm described above was implemented in Python, with a small portion that allows for the direct calculation of atomic orbitals in C. The one-electron integrals comprising the core Hamiltonian are calculated using a multigrid branch of PySCF.\citep{Sun} We benchmark the performance of the multigrid ISDF method with occ-RI using two systems: diamond, with a conventional unit cell containing eight carbon atoms, and lithium hydride, with a conventional unit cell containing four lithium and four hydrogen atoms. For all calculations the Goedecker-Teter-Hutter (GTH) pseudopotentials\citep{Goedecker96,Hartwigsen98,Hutter2019} and uncontracted GTH-CC-XZVP basis sets\citep{Hong-Zhou2022} of Ye are used throughout. We use a Kinetic energy cutoff of the plane waves in FFT of 70 $E_h$ for diamond and 130 $E_h$  for lithium hydride to ensure that the error from the finite plane-wave cutoff is less than 5 $\mu$Ha per atom. For all calculations shown in this section, we fix $\varepsilon_r=10^{-5}$ and $\alpha_{\min}=2.8$ Bohr$^{-2}$. 

\subsection{Accuracy of multigrid ISDF}
\begin{figure}[ht]
\centering\includegraphics[width=0.45\textwidth]{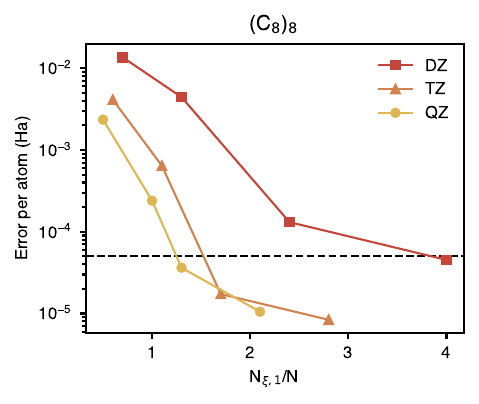}
\includegraphics[width=0.45\textwidth]{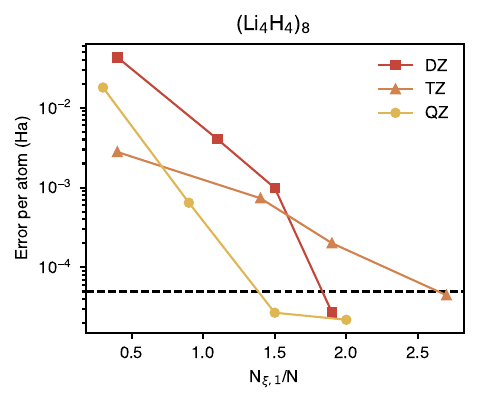}\\
\caption{
The error per atom incurred from the ISDF approximation for diamond $(\text{C}_8)$ and lithium hydride $(\text{Li}_4\text{H}_4)$ supercells with an increasing number of local fitting functions for fixed $\varepsilon_K$ of $10^{-2}$ and $10^{-3}$ for diamond and lithium hydride, respectively. The four points on each curve are obtained by using $\varepsilon_{ISDF}= 10^{-2}, 10^{-3}, 10^{-4}, 10^{-5}$. }\label{fig:locISDF}
\end{figure} 

In multigrid ISDF there are two types of fitting functions, the first set is local and is supported on a dense grid (their number is denoted by $N_{\xi,1}$). These functions are obtained through ISDF. The memory and CPU cost of using these functions increases quadratically with  $N_{\xi,1}$. The second type of ISDF functions are Sinc functions uniformly placed throughout the unit cell (their number is denoted by $N_{\xi,2}$). The memory and CPU cost of the calculation increases only linearly with the number of these functions. 

In Figure~\ref{fig:locISDF} we show how the number of local ISDF functions changes as one reduces the $\varepsilon_{ISDF}$ threshold and the accompanying reduction in the error of the calculation. As shown, the number of local ISDF functions needed is relatively small and the number of these functions does not change significantly with the size of the basis set. The error stops decreasing exponentially because we have kept $\varepsilon_K$ fixed which fixes the number of ISDF functions on the diffuse grid.

\begin{table}
    \centering
    \begin{tabular}{lcccccccccccc}
\hline
Basis&& $N_{\xi,1}$ && $N_{\xi,2}$ && $\alpha_1$ && $\alpha_2$ && $N$ && $N_{\mathrm{sharp}}$ \\
\hline
\multicolumn{13}{c}{Diamond $(\text{C}_4)_8$}\\
DZ &~~~~& 3240 &~~~&17576 &~~~& 4.3 -- 4.3 &~~~& 1.3 -- 0.1 &~~~& 1344 &~~~& 256\\
TZ &~~~~& 4128 &~~~~& 27000 &~~~~& 5.4 -- 5.4 &~~~~& 2.0 -- 0.1 &~~~~& 2368 &~~~~& 256\\
QZ &~~~~& 5256 &~~~~& 39304 &~~~~& 6.2 -- 6.2 &~~~~& 2.6 -- 0.1 &~~~~& 3968 &~~~~& 256\\
\\
\multicolumn{13}{c}{Lithium hydride $(\text{Li}_4\text{H}_4)_8$}\\
DZ &~~~~& 1728 &~~~&74088 &~~~& 8.4 -- 7.3 &~~~& 2.1 -- 0.1 &~~~& 896 &~~~& 160\\
TZ &~~~~& 4568 &~~~~& 39304 &~~~~& 10.9 -- 3.1 &~~~~& 1.4 -- 0.1 &~~~~& 1696 &~~~~& 320\\
QZ &~~~~& 6120 &~~~~& 74088 &~~~~& 12.5 -- 4.5 &~~~~& 2.3 -- 0.1 &~~~~& 3136 &~~~~& 320\\
\hline
\end{tabular}
    \caption{The basis functions and ISDF fitting functions for $2\times2\times2$ super cell of diamond ( (C$_8$)$_8$) and lithium hydride ( (Li$_4$H$_4$)$_8$). $N_{\xi,1}$: the number of ISDF functions on the dense grid, $N_{\xi,2}$: the number of uniformly placed Sinc functions on the diffuse grid, $\alpha_1$: the range of exponents local to the dense grid, $\alpha_2$: the range of exponents supported by the diffuse grid, $N$: the total number of basis functions, and $N_{\mathrm{sharp}}$: the number of sharp basis functions.}
    \label{tab:basISDF}
\end{table}

The number of uniform ISDF functions ($N_{\xi,2}$) depends on two settings: the largest GTO exponent that is less than the threshold $\alpha_{\min}$ and the threshold $\varepsilon_K$. These functions constitute the majority of the ISDF fitting functions and often far exceed the number of ISDF functions on the dense grid. $N_{\xi,2}$  can vary with the system and basis set. For example, in the TZ basis set of lithium hydride there are no exponents between 3.1 and 1.4 (see Table~\ref{tab:basISDF}), thus with the $\alpha_{\min}$ of 2.8 we find that the $N_{\xi,2}$ for the TZ basis set is smaller than that for the DZ basis set. Table~\ref{tab:basISDF} gives detailed information on the number of basis functions and ISDF functions supported on the dense and sparse grids, along with the range of exponents for the basis functions on each grid.

\begin{figure}[ht]
\centering
\includegraphics[width=0.45\textwidth]{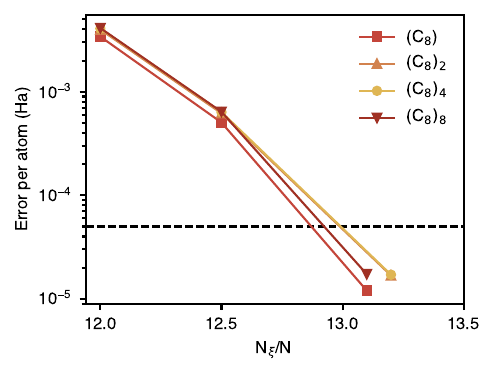}
\includegraphics[width=0.45\textwidth]{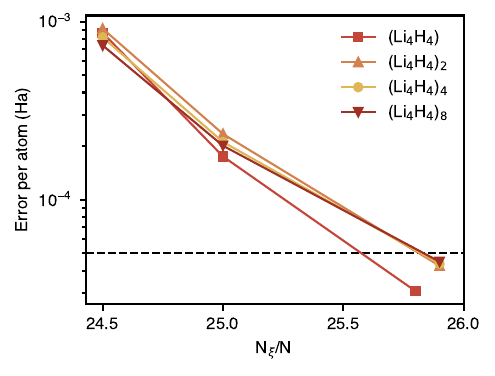}\\
\caption{ The results in the two graphs are obtained using the TZ basis set. $N_{\xi}$ in the graph includes ISDF basis functions from both the dense and sparse grids. The number of ISDF functions on the sparse grid was $N_{\xi,2}/N \approx 11$ for diamond, and 23 for lithium hydride. The accuracy is independent of system size so smaller systems may be used when choosing an ISDF threshold. The data here can be reproduced with the following settings: $\varepsilon_{K} = 10^{-2}$ and $\varepsilon_{ISDF} = 10^{-2}$, $10^{-3}$, and $10^{-4}$ for diamond, and $\varepsilon_{K} = 10^{-3}$ and $\varepsilon_{ISDF} = 10^{-3}$, $10^{-4}$, and $10^{-5}$ for lithium hydride.
}\label{fig:fullISDF}
\end{figure}

Next, we show that once the various thresholds are selected for a unit cell, the error per atom does not increase with the size of the system. Figure~\ref{fig:fullISDF} shows the error per atom with increasing $N_\xi/N$ for the diamond conventional unit cell $\text{C}_8$ and supercells of increasing sizes: $(\text{C}_8)_2$, $(\text{C}_8)_4$, and $(\text{C}_8)_8$; it also shows the lithium hydride conventional unit cell $\text{Li}_4\text{H}_4$ and supercells of increasing sizes: $(\text{Li}_4\text{H}_4)_2$, $(\text{Li}_4\text{H}_4)_4$, and $(\text{Li}_4\text{H}_4)_8$. Here the uncontracted GTH-CC-TZVP basis is used with $\varepsilon_{K} = 10^{-2}$ and $10^{-3}$ for diamond and lithium hydride, respectively, which determine the number of grid points in the sparse universal grid. Three different ISDF thresholds are used, $\varepsilon_{ISDF} = 10^{-2}, 10^{-3}, 10^{-4}$, which increase the number of ISDF fitting functions used on the dense grids. One can see that with decreasing thresholds the errors in the calculations exponentially decrease. The number of fitting functions $N_\xi$ includes the ISDF functions from the atom-centered grids and also all the grid points in the universal sparse grid. 

The $N_{\xi}/N$ in these calculations is larger than in conventional AO-based ISDF by almost a factor of two. It is worth remembering that the ISDF functions on the sparse grid most greatly contribute to $N_\xi/N$. For these functions, we do not store or calculate the two-center Coulomb integrals and thus, the overall CPU and memory cost is significantly lower than AO-based ISDF as we show in the next subsection.

\subsection{Cost of the multigrid calculations}
\begin{figure}[ht]
\centering
\includegraphics{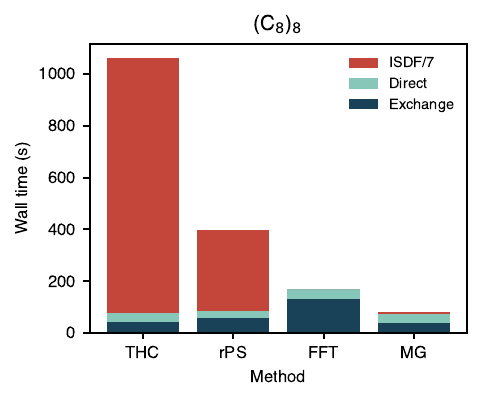}
\includegraphics{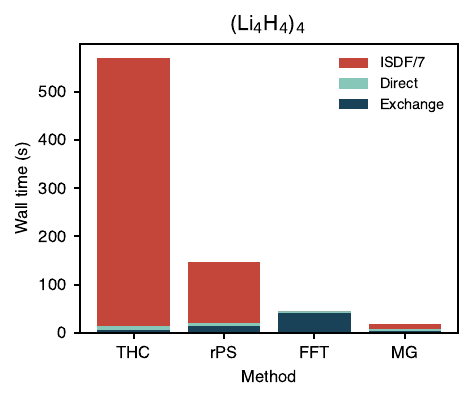}
\caption{Wall times for a single exchange build using the AO-based ISDF (THC), robust pseudospectral method (rPS), single-grid with no ISDF (FFT), and multigrid ISDF (MG) methods on a \emph{single core} of Intel(R) Xeon(R) CPU E5-2680 v3 @ 2.50GHz processor. The wall time for the full ISDF procedure (red) is divided by the number of iterations to evaluate the per-iteration cost. The systems reported are a 2$\times$2$\times$2 supercell of $\text{C}_8$ (diamond) and a 1$\times$2$\times$2 supercell of $\text{Li}_4\text{H}_4$ (lithium hydride). For an accuracy of $\approx$ 50 $\mu$Ha/atom, we used $\varepsilon_{K} = 10^{-2}$ and $\varepsilon_{ISDF} = 10^{-4}$ for diamond and $\varepsilon_{K} = 10^{-3}$ and $\varepsilon_{ISDF} = 10^{-5}$ for $\text{Li}_4\text{H}_4$ in the multigrid calculations. The ISDF parameters $N_\xi/N = $ 13 and 6 for $\text{Li}_4\text{H}_4$ and 7 and 4 for diamond were used for the THC and rPS methods, respectively, to get a similar accuracy.  For the MG method, the wall times for building the exchange (dark blue) and Coulomb (Direct, light blue) matrices are nearly equivalent. They are also comparable to the times for the THC and rPS methods. The ISDF wall time, however, is one to two orders of magnitude faster using the MG method. 
}\label{fgr:bar_chart}
\end{figure}

Figure~\ref{fgr:bar_chart} compares the wall times of single Coulomb and exchange builds and the ISDF wall time that includes the time to find ISDF functions $\xi$ and build the matrix $(\xi|\xi')$ (Eq.~\ref{eqn:(xi|xi')}). The ISDF times here are weighted by a factor of $1/7$. A major hurdle overcome by the multigrid method is the cost of ISDF. The wall times from the multigrid ISDF are one to two orders of magnitude faster than the single-grid THC and rPS methods (with occ-RI) from Ref.~\citep{Sharma2022}. In the multigrid ISDF method, this is reduced because only small subsets of product densities are fit on regions of the most dense grids (Eq.~\ref{eqn:M(R,R)}). The most significant efficiency gain comes from not fitting the universal sparse grid and calculating the matrix $(\xi|\xi')$ (Eq.~\ref{eqn:(xi|xi')}) in a direct fashion during the exchange build. 

The memory requirement for the largest arrays is significantly reduced in the multigrid builds. For the multigrid method, the matrix $(\xi|\xi')$ for the 2$\times2\times$2 diamond TZ data shown here is about 1 GB, and the orbital matrices $(\mu|\xi)$ require less than 1 GB.
For the LiH data shown here, the total memory required by the MG method is less than 2 GB. For diamond, the THC method requires about 53 GB and rPS 28 GB. For $1\times2\times2$ LiH supercell , THC and rPS require about 54 and 36 GB, respectively. Compared to the single grid methods, the MG method requires up to about 35 times less memory for the data here. The smaller system is used for LiH as the 2$\times$2$\times$2 supercell required more than 128 GB of memory when using the THC method.

\begin{figure}[ht]
\centering
\textbf{Diamond}\par\medskip
\includegraphics[width=0.3\textwidth]{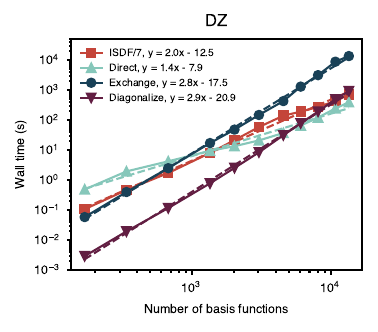}
\includegraphics[width=0.3\textwidth]{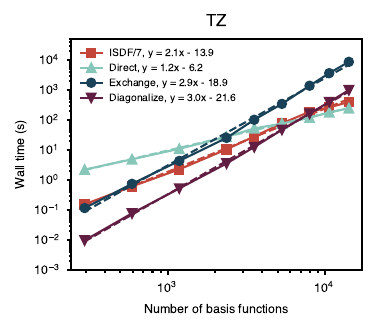}
\includegraphics[width=0.3\textwidth]{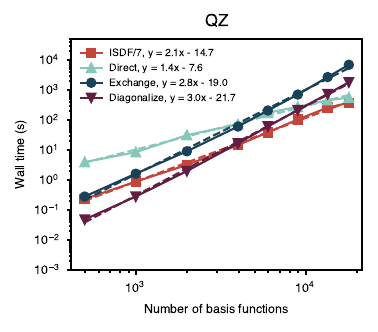}\\
\vspace{10mm}
\textbf{Lithium Hydride}\par\medskip
\includegraphics[width=0.3\textwidth]{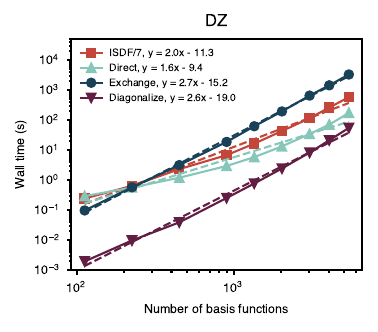}
\includegraphics[width=0.3\textwidth]{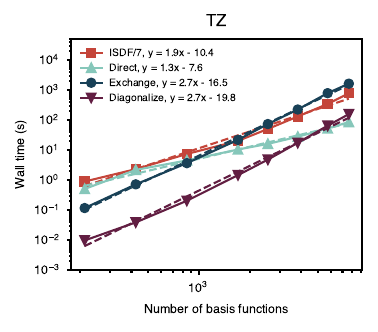}
\includegraphics[width=0.3\textwidth]{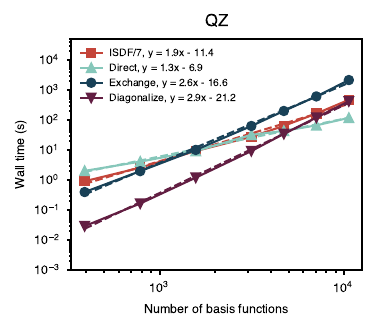}\\
\caption{Wall times for performing ISDF (weighted by 1/7, dark blue), a single build of the Coulomb (Direct, light blue) and exchange (purple) matrices, and diagonalizing the Fock matrix (red) on a \emph{single core} of Intel(R) Xeon(R) CPU E5-2680 v3 @ 2.50GHz processor. The calculations were performed using the uncontracted GTH-CC-DZVP (DZ, left), GTH-CC-TZVP (TZ, middle), and GTH-CC-QZVP (QZ, right) basis sets. The systems used are diamond supercells (top) of 1, 4, 8, 12, 18, 27, 36, 48, 64, and 80 unit cell copies (DZ), 1, 4, 8, 12, 18, 27, 36, and 48 copies (TZ), and 1, 4, 8, 12, 18, 27 and 36 copies (QZ), and lithium hydride supercells (bottom)  of 1, 4, 8, 12, 18, 27, 36, and 48 unit cell copies (DZ), 1, 4, 8, 12, 18, 27, and 36 copies (TZ), and 1, 4, 8, 12, 18, and 27 copies (QZ).  The settings used here are the same as in Figure~\ref{fgr:bar_chart}.
}\label{fgr:scaling}
\end{figure}

Next, we calculate the scaling of the various steps with increasing system size. Figure~\ref{fgr:scaling} shows the cost of single Coulomb and exchange builds, the total ISDF times (as in Fig.~\ref{fgr:bar_chart} are weighted by $1/7$), and the wall time per SCF iteration for diagonalizing the Fock matrix for diamond supercells of up to 80 unit cell copies, depending on the basis, and lithium hydride supercells of up to 48 copies. Calculations using GTH-CC-DZVP (DZ, left), GTH-CC-TZVP (TZ, middle), and GTH-CC-QZVP (QZ, right) are shown. The thresholds used are the same as reported in Fig.~\ref{fgr:bar_chart}. The following conclusions can be drawn from the graphs:
\begin{enumerate}
\item The multigrid ISDF scales quadratically $N^2$, which is due to the local nature of the fitting. For errors under 50 $\mu$Ha, we used an ISDF threshold $\varepsilon_{ISDF}$ of $10^{-4}$ for diamond and $10^{-5}$ for lithium hydride which corresponds to $N_\xi/N$ of 9 to 84, depending on the system and basis used. The majority of these fitting functions are Sinc functions on the sparse grid. A large value of 84 is sometimes needed for smaller basis set such as DZ because the number of Sinc functions can be quite large relative to the size of the basis set.
\item The scaling of the exchange matrix is cubic with the system size. The cost of the calculation for the largest basis set is only about a factor of 4 more expensive than the cost of performing diagonalization.
\item The cost of Coulomb is nearly linear with the size of the system. There is some deviation from linearity because in these calculations we only use 2-4 Coulomb grids. By adding more grids, linear scaling is possible;   doing so comes with additional CPU cost but no additional memory requirement.
\item The cost of the exchange is cheaper than the cost of Coulomb calculations for fairly large systems containing up to 3,000 basis functions in the case of diamond with the QZ basis set and for around 1,000 basis functions in the case of lithium hydride with the QZ basis set. 
\item The largest calculation, in terms of the number of basis functions,  was a $3\times 3\times 4$ diamond supercell with the QZ basis set containing 17,856 basis functions and 1,152 electrons. Because we use tighter thresholds, both $\varepsilon_{ISDF}$ and $\varepsilon_k$, for lithium hydride the largest calculation for it was a $3\times3\times3$ supercell containing 431 electrons and 10,584 basis functions; this was also with the QZ basis. The number of electrons per unit cell are also smaller in LiH than in Diamond. 
\end{enumerate}

\section{Conclusions}
In this paper, we have shown that efficient calculations of exchange matrices can be performed using our multigrid ISDF algorithm. The multigrid ISDF is significantly more efficient than the usual ISDF algorithm both in terms of memory and CPU cost. With this algorithm, relatively large calculations (containing $>17,000$ basis functions) can be performed without running out of memory on a single node. With this technique, the exchange calculation is more efficient than Coulomb calculations for systems with up to about 1,000 basis functions for TZ and QZ basis sets in diamond and lithium hydride systems. For large basis sets, such as QZ, the cost of exchange evaluation is only a factor of 4 more expensive than the cost of diagonalization and because the scaling of the two steps is similar we expect that this result will hold for even larger systems.

In our current implementation, we have not parallelized the calculations, although it should be possible to do so with high efficiency because most of the operations involve a series of FFTs that can be embarrassingly parallelized, or matrix multiplications that are also amenable to parallelization. We have also not made use of linear scaling approaches to leverage the fact that the exchange matrix is near-sighted, which can be used to further improve the efficiency of the calculations, particularly for systems with large band gaps. The approach we have outlined here has many extensions that we are actively exploring, including the ability to perform all-electron calculations, use of mixed Gaussian--plane-wave basis set, use for molecular calculations, the calculation of nuclear gradients, and also use of $k$-point symmetry at a nearly linear cost ($O(N_k \log(N_k)))$ with the number of k-points $N_k$. Given that the cost of the exchange evaluation is cheaper than Coulomb for systems containing up to 1,000 basis functions and around 100 electrons (these numbers are larger for diamond with the QZ basis set), one can likely perform hybrid DFT calculations with a similar cost as pure DFT calculations if the unit cells contain 100 or fewer electrons (because the scaling with $N_k$ is similar for both Coulomb and exchange).

\section{Acknowledgments}
K.E.S. was supported through the National
Science Foundation grant CHE-2145209 and S.S. through
a grant from the Camille and Henry Dreyfus Foundation.
\bibliographystyle{unsrt}


\end{document}